\documentclass[
twocolumn, 
aps, 
prapplied, 
superscriptaddress, 
amsmath, 
amssymb, 
]{revtex4-2}

\usepackage{changes}
\usepackage{graphicx}
\usepackage{dcolumn}
\usepackage{bm}
\usepackage{graphicx}                                
\usepackage{amssymb}
\usepackage{amsmath}
\usepackage{epsfig}
\usepackage{xcolor}
\usepackage{tabu}
\usepackage{mathtools}
\usepackage[colorlinks,linkcolor=blue,anchorcolor=blue,citecolor=blue,urlcolor=blue]{hyperref}
\usepackage{physics}
\usepackage{float}
\usepackage{diagbox}
\usepackage{inputenc}
\usepackage{soul}

\newcommand{\addrLab}{\affiliation{Laboratory of Quantum Information, University of Science and Technology of China, Hefei 230026, China}}

\newcommand{\addrSyn}{\affiliation{Synergetic Innovation Center of Quantum Information and Quantum Physics, University of Science and Technology of China, Hefei 230026, China}}

\newcommand{\addrSuzhou}{\affiliation{Suzhou Institute for Advanced Research, University of Science and Technology of China, Suzhou, Jiangsu 215123, China}}
\newcommand{\addrOrigin}{\affiliation{Origin Quantum, Hefei, Anhui 230088, China}}

\begin{document}

\title{Single-Step Phase-Engineered Pulse for Active Readout Cavity Reset with a Known Qubit State}

\author{Ren-Ze Zhao}
\addrLab\addrSyn
\author{Ze-An Zhao}
\addrLab\addrSyn
\author{Tian-Le Wang}
\addrLab\addrSyn
\author{Peng Wang}
\addrLab\addrSyn\addrSuzhou
\author{Sheng Zhang}
\addrLab\addrSyn\addrSuzhou
\author{Xiao-Yan Yang}
\author{Hai-Feng Zhang}
\author{Zhi-Fei Li}
\author{Yuan Wu}
\author{Sheng-Ri Liu}
\addrLab\addrSyn

\author{Peng Duan}
\email{pengduan@ustc.edu.cn}
\addrLab\addrSyn

\author{Guo-Ping Guo}
\email{gpguo@ustc.edu.cn}
\addrLab\addrSyn\addrOrigin

\date{\today}

\begin{abstract}
    
    In a circuit QED architecture, we experimentally demonstrate a hardware-efficient and  qubit-state-dependent Single-Step Phase-Engineered (SSPE) pulse scheme for actively depopulating a readout cavity. The protocol appends a reset segment with tailored amplitude and phase to a square readout pulse. Within the linear-response regime, the optimal reset amplitude scales proportionally with the readout amplitude, while the optimal reset phase remains invariant, significantly simplifying the experimental calibration procedure. For a known qubit state, time-resolved measurements demonstrate that the SSPE pulse achieves faster cavity depletion than the CLEAR protocol. Crucially, this approach enables arbitrarily fast, overshoot-free depletion of the cavity photon population, with the ultimate reset rate constrained by the available drive amplitude and the finite analog bandwidth of the measurement chain. A comprehensive evaluation of the QND character reveals no measurable increase in non-QNDness relative to the free-decay and CLEAR protocols. The remaining non-QNDness is predominantly attributable to state switching induced by qubit relaxation during the readout process. These results establish the SSPE scheme as a practical and scalable approach for achieving rapid and smooth cavity reset in superconducting quantum circuits in the case of a known qubit state. 
\end{abstract}

\maketitle

\section{Introduction}

Superconducting qubits realized through circuit quantum electrodynamics~\cite{2021cqed_review, 2004cqed, 2007cqed, 2019_engineer_review} (QED) have emerged as a leading platform for scalable quantum computation. This architecture has enabled the implementation of high-fidelity two-qubit gates~\cite{2024_high_fidelity_CZ_double_transmon_coupler, 2025_IQM_high_Single_Two_Readout}, fast and accurate single-shot qubit state readout~\cite{2023_2state_readout, 2025_IQM_high_Single_Two_Readout, 2024_dynamic_readout, 2025_MTL_filter, 2025_Pan_longitudinal_readout, 2025_tunable_purcell_filter_YuDapeng}, and, notably, the experimental demonstrations of quantum error-correction codes have surpassed the fault-tolerance threshold~\cite{2023_surface_code_Google, 2025_surface_code_Google, 2025_color_code_google, 2025_dynamic_code_google, 2023_bosonic_code, 2023_devoret_qec}, marking an essential milestone toward fault-tolerant quantum computing.

A key component of the circuit QED readout scheme is dispersive measurement, wherein the qubit is coupled to a readout resonator cavity. In the dispersive regime, the qubit state induces a frequency shift of the resonator without energy exchange. When the resonator is driven near resonance, its vacuum state is displaced into a qubit-state-dependent coherent state. By demodulating the transmitted or reflected microwave signal, one can infer the qubit state with high fidelity.

Rapid removal of residual photons from a readout resonator is important whenever subsequent qubit operations must begin before the cavity field has completely decayed to vacuum, because residual photons increase the measurement dead time and can induce photon-number-dependent ac-Stark shifts and dephasing~\cite{ac_stark_shift, number_spiltting, photon_dependent_dephasing}. These considerations underscore the importance of active and high-performance cavity depopulation techniques in modern superconducting quantum computation.

A straightforward strategy to accelerate resonator reset is to engineer a low–quality-factor readout resonator. However, increasing the resonator linewidth enhances the qubit’s radiative decay through the Purcell effect. Purcell filters can mitigate this decay channel, though they generally introduce additional circuit complexity, such as dedicated filter cavities~\cite{dedicated_filter, 2025_MTL_filter} or bus filter structures~\cite{passband_bus_filter, stopband_bus_filter}. Furthermore, for a fixed ratio of the dispersive shift to resonator linewidth as desired for a high-fidelity readout~\cite{quantum_jump}, increasing linewidth also enhances qubit dephasing induced by stray photons~\cite{photon_noise1, photon_noise2, photon_noise3}. Parametric driving offers an alternative approach to rapid cavity reset, albeit at the cost of integrating an auxiliary dissipative mode into the circuit~\cite{parameteric_drive_cavity_reset}. 

Pulse-level approaches provide a complementary route to fast cavity depletion without modifying the hardware architecture. Optimal control techniques have also been applied to construct tailored drive sequences that efficiently deplete cavity photons~\cite{optimal_control_cavity_reset}. In the nonlinear dispersive regime, although optimization algorithms have been employed to derive drive parameters for both conditional and unconditional resonator reset, the resulting reset durations remain long compared to typical gate times~\cite{nonlinear_cavity_reset}. Two additional segments are appended to the end of the pulse as a counter pulse to achieve cavity reset, as demonstrated in Ref.~\cite{CLEAR}.

\begin{figure}
    \centering
    \includegraphics[width=\columnwidth]{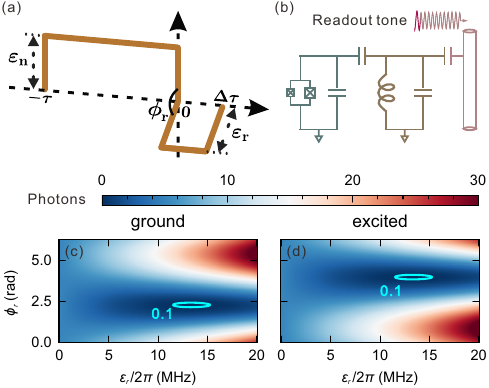}
    \caption{\textbf{Structure and Performance of the SSPE Pulse.} \textbf{(a)} Schematic of the SSPE pulse, which consists of two successive segments: a normal readout segment with drive amplitude $\varepsilon_n$ and phase $\phi_n$, followed by a reset segment with drive amplitude $\varepsilon_r$ and phase $\phi_r$. \textbf{(b)} Equivalent quasi-lumped-element circuit model. \textbf{(c), (d)} Simulated residual cavity photon number at the end of the SSPE pulse as a function of the reset amplitude and phase, for the qubit prepared in $\ket{g}$ and $\ket{e}$, respectively. A residual photon number of zero indicates that the cavity field has returned to vacuum. Light cyan contours indicate the photon number of 0.1. Qubit parameters used in the simulation are summarized in Table~\ref{tab:Dev_params}.}
    \label{fig-fig1}
\end{figure}

In this letter, we present an experimental demonstration of a qubit-state-dependent Single-Step Phase-Engineered (SSPE) active cavity reset technique that enables smooth and highly efficient photon depletion. As illustrated in Fig.~\ref{fig-fig1}{(a)}, a reset segment with tailored amplitude and phase is appended to the trailing edge of a square readout pulse. Within the linear-response regime, the SSPE protocol exhibits a linear-amplitude and fixed-phase (LAFP) scaling property, a feature that drastically simplifies the experimental calibration procedure. For a known prepared qubit state with correspondingly selected reset parameters, the time-resolved cavity photon number measurements demonstrate faster cavity depletion with SSPE than the CLEAR~\cite{CLEAR}. It facilitates arbitrarily fast photon depletion without introducing transient overshoots, bounded by the maximum achievable drive amplitude of the hardware and the finite analog bandwidth of the measurement chain. We further characterize the quantum non-demolition (QND) properties of the readout operation, our results demonstrate that the SSPE pulse introduces no additional non-QND measurement errors (non-QNDness), exhibiting a residual non-QNDness that is completely comparable to both the free-decay and the CLEAR protocols. This residual non-QNDness is found to be predominantly governed by state switching induced by intrinsic qubit relaxation during the readout process.

\section{Theoretical Model}

In a circuit QED architecture, a superconducting qubit is coupled to a superconducting resonator cavity, typically implemented as a quarter-wavelength ($\lambda/4$) coplanar waveguide transmission line~\cite{cpw}. This system can be modeled by an equivalent quasi-lumped-element circuit in which the qubit is capacitively coupled to a linear LC oscillator, as illustrated schematically in Fig.~\ref{fig-fig1}(b). When a microwave pulse with a frequency near resonance is injected into the input port, the cavity field evolves into a coherent field. The dynamics of the coherent field amplitude $\alpha(t)$ can be obtained from input–output theory~\cite{input-output_theory1, input-output_theory2, input-output_theory3}, and can be shaped through appropriate engineering of the drive envelope. The resulting dynamics are governed by the differential equation~\cite{2021cqed_review, number_splitting}

\begin{equation}
    \Dot{\alpha}_j (t) = -i \varepsilon_d (t) - i (\Delta_r + \chi_j) \alpha_j(t) - \frac{\kappa}{2} \alpha_j(t),
    \label{eq-1}
\end{equation}
where $\Delta_r = \omega_r - \omega_d$ denotes the detuning between the bare cavity frequency $\omega_r$ and the cavity drive frequency $\omega_d$. Here, $\varepsilon_d(t)$ represents the (generally complex) time-dependent amplitude of the external drive, and $\kappa$ is the cavity decay rate. The quantity $\chi_j$ denotes the dispersive shift of the cavity resonance frequency associated with the qubit being in state $\ket{j}$, where $j = 0, 1, \dots$ correspond to the ground, first excited, and higher excited states, respectively. For $j \textgreater 0$, we have

\begin{align}
    \chi_j &= \chi_{j-1,j} - \chi_{j,j+1} , \nonumber \\
    \chi_{j-1,j} & = \frac{j g^2}{\Delta + (j-1) \eta},
\end{align}
In particular, $\chi_0=-g^2/\Delta$, where $g$ is the qubit-cavity coupling strength, $\eta$ is the qubit anharmonicity, and $\Delta=\omega_q - \omega_r$ is the detuning between the qubit frequency $\omega_q$ and the bare cavity frequency. In this work, we restrict our analysis to the cavity reset dynamics for the qubit prepared in either $\ket{0}$ or $\ket{1}$ (equivalently, $\ket{g}$ or $\ket{e}$). The instantaneous mean cavity photon number is given by

\begin{equation}
    \bar{n}_j(t) = | \alpha_j (t) | ^ 2.
\end{equation}

We emphasize that our SSPE pulse is constructed from piecewise-constant complex amplitudes and consists of two segments: a normal readout segment of duration $\tau$ with drive amplitude $\varepsilon_n$ and phase $\phi_n$, followed by a reset segment of duration $\Delta \tau$ with amplitude $\varepsilon_r$ and phase $\phi_r$.
\begin{align}
		\varepsilon_d(t) =
		\begin{cases}
			\varepsilon_n e^{i \phi_n}, & -\tau \le t < 0, \\
			\\ 
			\varepsilon_r e^{i \phi_r}, & 0 \le t \le \Delta \tau,
		\end{cases}
\end{align}
Throughout this work, both $\varepsilon_n$ and $\varepsilon_r$ are taken to be real. The relative phase $\Delta \phi = \phi_r - \phi_n$ determines the dynamics of the cavity coherent field and plays a crucial role in determining the reset efficiency.

Under the drive of this piecewise-constant pulse, the coherent field amplitude $\alpha_j(t) $ evolves accordingly

\begin{align}
    \alpha_j(t) =
    \begin{cases}
        \frac{2 \varepsilon_n \left(-1 + e^{ -\frac{1}{2} (t+\tau) C_j }\right) }{-iC_j}, & -\tau \le t < 0, \\
        \\
        2 e^{-\frac{1}{2} t C_j} \\
        \times \frac{\varepsilon_r e^{i\phi_r} \left(1-e^{\frac{1}{2} t C_j}\right) - \varepsilon_n \left(1-e^{-\frac{1}{2} \tau C_j}\right)}{-iC_j}, & 0 \le t \le \Delta \tau,
    \end{cases}
    \label{eq-5}
\end{align}
where $C_j = 2i\Delta_r + \kappa + 2i\chi_j$. For a fixed readout amplitude $\varepsilon_n$ and duration $\tau$, typically optimized for high readout fidelity, and for a given qubit state and a chosen reset duration $\Delta \tau$, there always exists a pair of reset parameters $\left( \varepsilon_r,  \phi_r \right)$ that returns the cavity field to the vacuum state at the end of the pulse, i.e.

\begin{equation}
    {\left| \alpha_j(\Delta \tau) \right|}^2 = 0.
\end{equation}

Owing to the state-dependent dispersive shift $\chi_j$, the optimal complex reset parameter is determined separately for $\ket{g}$ and $\ket{e}$, as derived in Appendix~\ref{sec: appendixD} and demonstrated in Fig.~\ref{fig-fig1}(c) and \ref{fig-fig1}(d), respectively. In principle, for sufficiently large drive amplitude, cavity reset can be performed in an arbitrarily short time. In practice, however, the achievable depletion rate is limited by both the maximum achievable drive amplitude of the hardware and the finite analog bandwidth of the measurement chain. The minimum achievable reset duration supported by our system is discussed in Appendix~\ref{sec: appendixB}. We further highlight a key feature of our scheme: upon rescaling the readout amplitude, the optimal reset amplitude scales proportionally, whereas the optimal reset phase remains invariant. A detailed derivation of this result is provided in Appendix~\ref{sec: appendixD}. This scaling property substantially simplifies the experimental calibration procedure and will be further demonstrated in the following sections.

\section{Experimental Results}

The experimental qubit is a frequency-tunable transmon~\cite{transmon} with frequency $\omega_q / 2\pi = 5350$~MHz at the sweet spot, anharmonicity $\eta / 2\pi=-218$~MHz, average $T_1 = 113$~$\rm{\mu s}$. The bare cavity resonance frequency is measured to be $\omega_{\text{bare}} / 2\pi=7102.5$~MHz. When the transmon is prepared in $\ket{g}$, the dressed cavity frequency shifts to $\omega_{\ket{g}} / 2\pi=7107.9$~MHz. Preparing it in $\ket{e}$ results in a dispersive shift of $2\chi / 2\pi = -2.2$~MHz. The cavity decay rate is $\kappa / 2 \pi = 2.4$~MHz, corresponding to a photon lifetime of $T_{\text{cav}} = 1 / \kappa \approx 66$~ns. In the dispersive regime, the critical photon number is approximately given by $n_\text{crit}=\Delta^2 / 4 g^2\approx 44$~\cite{2021cqed_review}. 
Throughout this work, the cavity drive frequency is tuned to the midpoint between the dressed cavity resonance frequencies associated with the qubit in $|g\rangle$ and $|e\rangle$. This choice maximizes the phase response contrast and ensures identical steady state photon populations for both qubit states, provided that the cavity nonlinearities remain negligible. Further details of the experimental setup are provided in Appendix~\ref{sec: appendixA}.

\begin{figure}
    \centering
    \includegraphics[width=\columnwidth]{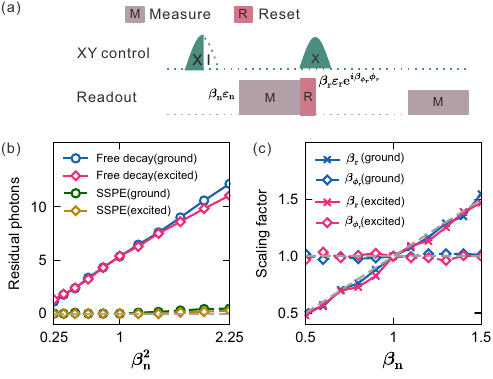}
    \caption{\textbf{Experimental Pulse Sequence and the LAFP Property  of the SSPE Pulse.} \textbf{(a)} Experimental pulse sequence used to characterize cavity photon number. The qubit is prepared in either $\ket{g}$ or $\ket{e}$, followed by the primary readout pulse and its appended reset segment. To measure the residual photon population, a frequency-swept $\pi$-pulse probes the photon-induced ac-Stark shift. The midpoint of this spectroscopic $\pi$-pulse is precisely aligned with the trailing edge of the reset segment. A subsequent square readout pulse is then applied for high-fidelity qubit-state discrimination. The delay between the primary and the subsequent readout pulses is set to 500~ns, ensuring the complete depletion of residual cavity photons prior to the second measurement. For the primary readout, the amplitude is scaled as $\beta_n \varepsilon_n$, and the corresponding reset segment is expressed as $\beta_r \varepsilon_r e^{i \beta_{\phi_r} \phi_r}$. Both the reset segment and the spectroscopic $\pi$-pulse have a fixed duration of 30~ns. For the Square configuration, the reset segment is replaced by a passive free-decay interval. \textbf{(b)} Residual photon population as a function of the squared drive amplitude, parameterized by the scale factor $\beta_n$, where $\beta_n=1$ corresponds to the calibrated readout amplitude yielding high readout fidelity. \textbf{(c)} Experimental verification of the LAFP scaling property of the SSPE scheme. The required scaling factors for the optimal reset amplitude ($\beta_r$, crosses) and the optimal reset phase ($\beta_{\phi_r}$, open diamonds) are plotted against the readout amplitude scaling factor $\beta_n$. Data points corresponding to the qubit ground and excited states are indicated by blue and pink symbols, respectively. Gray dashed lines indicate the reference lines $y=1$ and $y=x$.}
    \label{fig-fig2}
\end{figure}

To characterize the reset performance, we first fix the duration of the reset segment to $\Delta \tau = 30$~ns $\approx T_{\text{cav}} / 2$. After determining the optimal reset parameters $\left( \varepsilon_r,  \phi_r \right)$, we quantify the performance of the cavity reset by measuring the residual photon number at a time after the readout pulse. To extract this quantity, we employ the pulse sequence illustrated in Fig.~\ref{fig-fig2}(a). A calibrated qubit $\pi$-pulse is frequency-swept to measure the photon-induced ac-Stark shift~\cite{ac_stark_shift}, $\Delta \omega_q = 2 \chi n$, where $n$ denotes the mean cavity photon number. To experimentally confirm the LAFP scaling property of the SSPE scheme, we additionally rescale the amplitude of the readout segment by a dimensionless scaling factor $\beta_n$.

For a fair comparison, the residual photon population following the SSPE pulse is compared against that of the Square pulse supplemented by an extra passive free-decay delay of duration $\Delta \tau$, ensuring that both schemes have equal total duration. The measured results are shown in Fig.~\ref{fig-fig2}(b). For all readout segment amplitudes, parameterized by the scaling factor $\beta_n = \varepsilon_n^{'}/\varepsilon_n$, the SSPE pulse consistently suppresses the residual photon population to zero, demonstrating highly efficient cavity depopulation. The slight increase in residual photon number observed at large drive amplitudes is attributed to temporal averaging effects induced by the finite $\pi$-pulse duration, which become more pronounced at elevated cavity photon numbers. In contrast, the free-decay scheme exhibits a residual photon population that increases linearly with the square of drive amplitude, consistent with the cavity’s linear-response behavior. At sufficiently large drive amplitudes, however, nonlinear effects become appreciable~\cite{cavity_nonlinearity_dressed_dephasing_model, cavity_nonlinearity_improve_readout, cavity_nonlinearity_jc_nonlinearity, mist_transmon_ionization, mist_transmon_ionization_dynamics}: the cavity frequency acquires a photon-number-dependent shift, giving rise to qubit-state-dependent photon populations, as evident in the large amplitude regime of Fig.~\ref{fig-fig2}(b). The robustness of the SSPE pulse against these nonlinearities underscores its suitability for operation in both the linear and weakly nonlinear regimes.

We further highlight that the SSPE reset scheme exhibits a practically convenient amplitude-scaling property. When the readout segment amplitude is rescaled by a factor $\beta_n$, the corresponding optimal reset amplitude rescales proportionally as $\varepsilon_r^{\prime} = \beta_r \varepsilon_r$, with $\beta_r = \beta_n$. In contrast, the optimal reset phase is invariant under this rescaling, i.e., $\beta_{\phi_r} = \phi_r^{\prime} / \phi_r = 1$. This LAFP scaling property, summarized in Fig.~\ref{fig-fig2}(c), substantially simplifies the experimental calibration procedure: once the optimal reset phase is determined, it can be retained as the readout amplitude is varied within the linear-response range, while the corresponding reset amplitude can be estimated by proportional rescaling. 

To probe the cavity photon number dynamics during the reset stage, we employ the pulse sequence depicted in Fig.~\ref{fig-fig2}(a), with the delay of the spectroscopic $\pi$-pulse varied systematically to obtain time-resolved measurements of $n(t)$. The efficiency of the SSPE scheme is benchmarked against the free-decay and CLEAR protocols. For a consistent and fair comparison, the total reset segment durations of the SSPE and CLEAR pulses are set to be equal, with the two individual segments of the CLEAR protocol constrained to be of equal duration.

\begin{figure}
    \centering
    \includegraphics[width=\columnwidth]{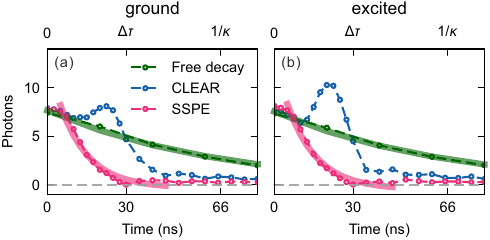}
    \caption{\textbf{Cavity Photon Number Dynamics.} \textbf{(a), (b)} Time-resolved cavity photon number dynamics during the reset stage for all three reset protocols, with the qubit initialized in $\ket{g}$ or $\ket{e}$, respectively. The reset segment duration is uniformly set to 30~ns for both the SSPE (pink) and CLEAR (blue) protocols. Solid green and pink curves denote exponential fits to the passive free decay and SSPE reset trajectories, respectively. Complete time-resolved photon number dynamics, including data for incorrectly initialized qubit states, are provided in Appendix~\ref{sec: appendixB}.}
    \label{fig-fig3}
\end{figure}

The measured cavity photon number dynamics for all three reset protocols with the qubit initialized in $\ket{g}$ and $\ket{e}$, are shown in Fig.~\ref{fig-fig3}(a) and~\ref{fig-fig3}(b) respectively. For the free-decay protocol, the photon number follows a standard exponential decay, with fitted decay rates of $\kappa_{\ket{g}} / 2 \pi = 2.37$~MHz and $\kappa_{\ket{e}} / 2 \pi  = 2.39$~MHz.  In contrast, the SSPE pulse drives the cavity population rapidly and smoothly to zero, with exponential fits yielding effective decay rates of $\kappa_{\ket{g}}^{\text{SSPE}} / 2 \pi  = 14.44$~MHz and $\kappa_{\ket{e}}^{\text{SSPE}}  / 2 \pi = 14.12$~MHz, approximately six times faster than the passive decay rate. The CLEAR protocol, however, fails to achieve complete cavity depletion within the allotted reset duration, and exhibits a pronounced photon number overshoot prior to depletion. For our SSPE scheme, the achievable depletion rate is limited by both the maximum achievable drive amplitude of the hardware and the finite analog bandwidth of the measurement chain, the latter of which inevitably distorts the rapidly varying transient signals; experimentally, we observe performance degradation for reset durations below 10~ns. The CLEAR protocol is subject to an inherent trade-off between depletion rate and suppression of the cavity photon number overshoot. The CLEAR pulse capable of achieving complete cavity photon depletion at a reset duration of 100~ns is presented in Appendix~\ref{sec: appendixB}.

 The measured output field IQ trajectories are consistent with these observations. Under the SSPE protocol, both quadrature components converge to the origin at the termination of the pulse, confirming complete cavity depletion. Under the CLEAR protocol, the IQ components decay to a finite nonzero value by the end of the pulse, indicative of incomplete photon depletion. A detailed theoretical derivation of the three-port input-output relations, together with calibration details and additional output field IQ trajectory data, is provided in Appendix~\ref{sec: appendixC}.

\begin{figure}
    \centering
    \includegraphics[width=\columnwidth]{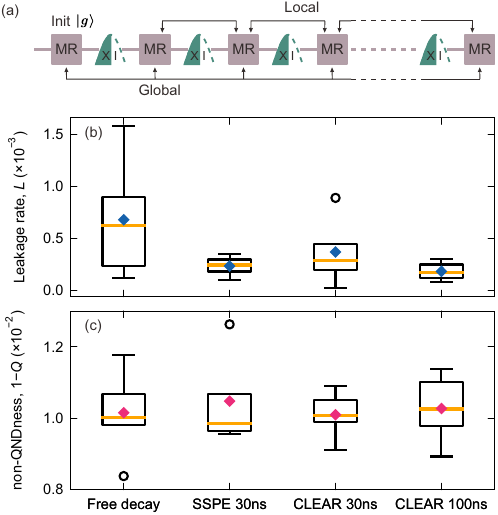}
    \caption{\textbf{QNDness Characterization.} \textbf{(a)} Pulse sequence for the RILRB experiment. Measurement operations are interleaved with randomized qubit flips. The local and global correlations are illustrated by the upper and lower arrows, respectively. The block labeled ``MR" denotes the combined measurement and reset components. \textbf{(b)} leakage rate and \textbf{(c)} non-QNDness for different pulse configurations, shown as box plots. Each box extends from the first to the third quartile of the data, with the horizontal lines indicating the median and the diamond markers indicating the mean. The whiskers extend to the farthest data points within 1.5 times the inter-quartile range from the box; points beyond the whiskers are shown as fliers.}
    \label{fig-fig4}
\end{figure}

It is essential to quantify the QND nature of the readout pulse. Readout-induced qubit leakage~\cite{mist_beyond_rwa, mist_within_rwa, mist_transmon_ionization, mist_transmon_ionization_dynamics, mist_dispersive_readout, mist_four-wave_mixing, mist_spectroscopy, mist_LZ} and relaxation~\cite{mist_t1_anti-zeno, mist_t1_1, mist_t1_2, mist_t1_3, mist_t1_4} are well documented in dispersive measurements of superconducting qubits. We characterize these effects using the readout-induced leakage randomized benchmarking (RILRB) protocol introduced in Ref.~\cite{rilrb_prl} and adapted in Ref.~\cite{2025_IQM_high_Single_Two_Readout}. As depicted in Fig.~\ref{fig-fig4}(a), the pulse sequence begins with a pre-measurement SSPE reset pulse, and subsequent measurement operations are interleaved with randomized qubit flips. Each measurement operation comprises a readout component and a reset component. The reset component lasts for 400~ns, ensuring that the cavity photons fully dissipate before the subsequent measurement, even in the presence of erroneous state preparation in the SSPE scheme during the sequence. Within an initial time window $\Delta \tau$ of this reset component, various active reset protocols are applied. The readout duration is fixed at 900~ns, yielding a total cycle time of $t_{\text{cycle}} = 1.3\ \mu\text{s}$. The local correlations quantify the correlations between two consecutive readout outcomes, whereas the global correlations quantify the correlations between the initial state and the $n$-th readout outcome. See Appendix~\ref{sec: appendixE} for more details on the experimental protocol. Following Ref.~\cite{2025_IQM_high_Single_Two_Readout}, the local and global correlations are modeled as

\begin{equation}
    {\langle \bar{C}_n^{\text{local}}\rangle} = J \left( 1-L-S \right)^n + K,
\end{equation}

\begin{equation}
    {\langle \bar{C}_n^{\text{global}}\rangle} = M \left( 1 - 2p - L \right)^n + N,
\end{equation}
where $L$ and $S$ denote the average leakage and seepage rate, respectively, and $p$ is average state switch probability per readout. The coefficients $J,\ K,\ M\ \text{and}\ N$ account for readout error, state-initialization error, and Pauli errors accumulated during the sequence. Further details of the models are provided in Appendix~\ref{sec: appendixE}.
The QND fidelity $Q$ is evaluated by simultaneously fitting both the correlation functions, yielding

\begin{equation}
    Q = 1-p-L,
\end{equation} 

The extracted leakage rates and non-QNDness for various pulse configurations are shown in Fig.~\ref{fig-fig4}(b) and~\ref{fig-fig4}(c). The tested configurations include free decay, a 30-ns SSPE sequence, and CLEAR protocols with durations of 30~ns and 100~ns, corresponding respectively to incomplete and complete cavity photon depletion. All tested schemes exhibit negligible leakage rates ($L \approx 10^{-4}$), as indicated by the nearly flat local correlation curves; representative raw fitting data are provided in Appendix~\ref{sec: appendixE}. The measured non-QNDness is therefore dominated almost entirely by the state-switching rate. As shown in Fig.~\ref{fig-fig4}(c), no discernible differences in non-QNDness are observed among the tested reset schemes. We attribute the observed state-switching rate primarily to qubit relaxation during the readout process rather than to the reset protocol itself. The estimated qubit relaxation probability per cycle, $1-e^{-t_{\text{cycle}}/T_1} \approx 1.23\%$, is comparable to the measured state-switching rate. These results demonstrate that the SSPE protocol achieves rapid depletion without introducing any additional non-QND measurement errors.

\section{Conclusion and Outlook}

In summary, we have demonstrated an efficient SSPE protocol for active cavity reset in circuit QED systems. By appending a reset segment with tailored amplitude and phase to a square readout pulse, the scheme achieves rapid and smooth cavity photon depletion without requiring any modification to the hardware architecture. Within the linear-response regime, the LAFP scaling property allows the optimal reset phase to be retained as the readout amplitude is varied, while the corresponding reset amplitude can be estimated by proportional rescaling. This property substantially reduces the experimental calibration overhead. Time-resolved measurements of the cavity photon number demonstrate that SSPE achieves faster cavity depletion than CLEAR when the qubit is prepared in a known state. The protocol facilitates arbitrarily rapid photon evacuation entirely devoid of transient overshoots, establishing that its operational speed limit is governed by the maximum achievable drive amplitude of the hardware and the finite analog bandwidth of the measurement chain. Comparative benchmarking against both free-decay and CLEAR schemes demonstrates that the SSPE protocol introduces no additional non-QND measurement errors, with the residual non-QNDness dominantly governed by state switching induced by intrinsic qubit relaxation during the readout process.

A direct application of the SSPE scheme is heralded state initialization via post-selection~\cite{heralded_state_initialization}. In this setting, the reset pulse is calibrated for the desired qubit state, and experimental shots in which the heralding measurement identifies an undesired state are discarded. Future work will explore the design of time-varying amplitude and phase envelopes of the reset segment to realize qubit-state-independent cavity reset, and will extend the method to multi-qubit architectures that share a common readout resonator~\cite{2009_two_qubit_readoout, 2009_two_qubit_readoout_nature, 2023_two_qubit_readoout}.

\begin{acknowledgments}
This work is supported by the National Natural Science Foundation of China (Grant No.~12034018). This work is also partially carried out at the USTC Center for Micro and Nanoscale Research and Fabrication. 
\end{acknowledgments}

\section*{data availability}
The data that support the findings of this article are not
 publicly available. The data are available from the authors
 upon reasonable request.

\appendix
\counterwithout{figure}{section}
\counterwithout{table}{section}
\numberwithin{equation}{section}
\renewcommand{\thesection}{\Alph{section}}
\renewcommand{\thetable}{\Roman{table}}

\section{Experimental Setup}
\label{sec: appendixA}

Our experiments are performed on an uncoupled superconducting quantum processor comprising two transmon qubits. Both qubits employ asymmetric superconducting quantum interference device (SQUID) loops, enabling frequency tunability via externally applied magnetic flux. Each qubit is equipped with dedicated control and measurement hardware, including an XYZ control line for single-qubit rotations and flux tuning, and an individual dispersive readout resonator. A notch-type Purcell filter—comprising two $\lambda/4$ resonators operating in reflection—is coupled to the bus line. An input capacitor $C_{\text{in}}$ provides directional signal routing, preferentially directing the readout signal toward the output port to minimize loss into the input port. The relevant qubit parameters are summarized in Table~\ref{tab:Dev_params}.

The quantum processor is installed at the mixing chamber (MC) stage of a dilution refrigerator operating at a base temperature of approximately 20~mK. Signal delivery and readout are facilitated by a cryogenic wiring network incorporating multiple stages of attenuation and filtering to suppress thermal and technical noise. Flux control is generated using an arbitrary waveform generator (AWG) with a sampling rate of 1.2~GSa/s, enabling both static flux biasing and fast flux-pulse operation. Qubit drive and readout signals are synthesized using a radio-frequency (RF) module covering 3.75–8.25~GHz with a sampling rate of 3.2~GSa/s. Dispersive readout is performed in transmission: the readout signals produced by mixing DAC outputs with a local-oscillator (LO) tone propagate through an impedance-matched parametric amplifier (IMPA)~\cite{IMPA} located at the MC stage, followed by a high-electron-mobility transistor (HEMT) amplifier at the 4~K stage. Subsequent room-temperature amplification and downconversion to an intermediate frequency (IF) precede digitization and demodulation, after which the extracted in-phase (I) and quadrature (Q) components are used for high-fidelity quantum-state discrimination.

\begin{table*}
    \renewcommand\tabcolsep{36pt}
    \centering
    \caption{Device parameters} 
    \begin{tabular}{c|c}
    \hline
    \hline
    \\
    Device Parameters & $Q_1$ \\
    \\
    \hline
    Qubit 0-1 transition frequency at the sweet spot $\omega_q/ 2\pi$~(MHz)             & 5350 \\
    Qubit anharmonicity $\eta / 2\pi$~(MHz)                                             & -218  \\
    $T_1$ at the sweet spot~($\mu$s)                                                    & 113  \\
    $T_1$ at photon-induced ac-Stark-shifted frequency~($\mu$s)                         & 111  \\
    $T_2^{\ast}$ at the sweet spot~($\mu$s)                                             & 118  \\
    $T_2^\text{echo}$ at the sweet spot~($\mu$s)                                        & 122  \\
    \hline
    Bare cavity resonance frequency $\omega_\text{bare} / 2\pi$~(MHz)                   & 7102.5 \\
    Cavity resonance frequency when qubit in $\ket{g}$ $\omega_{\ket{g}} / 2\pi$~(MHz)  & 7107.9  \\
    Double dispersive shift $2\chi / 2 \pi$~(MHz)                                       & -2.2  \\
    Cavity decay rate when qubit in  $\ket{g}\ \kappa_{\ket{g}} / 2 \pi$~(MHz)          & 2.37  \\
    Cavity decay rate when qubit in  $\ket{e}\ \kappa_{\ket{e}} / 2 \pi$~(MHz)          & 2.39 \\
    Total readout pulse duration~(ns)                                                   & 900 \\
    Qubit-Cavity coupling strength $g / 2 \pi$~(MHz)                                    & 132  \\
    Dispersive critical photon number $n_\text{crit}$                                   & 44  \\
    Average readout fidelity of $\ket{g}\ F_{g}$ ($\%$)                                        & 99.4  \\
    Average readout fidelity of $\ket{e}\ F_{e}$ ($\%$)                                        & 97.7  \\
    \hline
    \hline
    \end{tabular}
    \label{tab:Dev_params}
\end{table*}

\begin{figure*}
    \includegraphics[width=2\columnwidth]{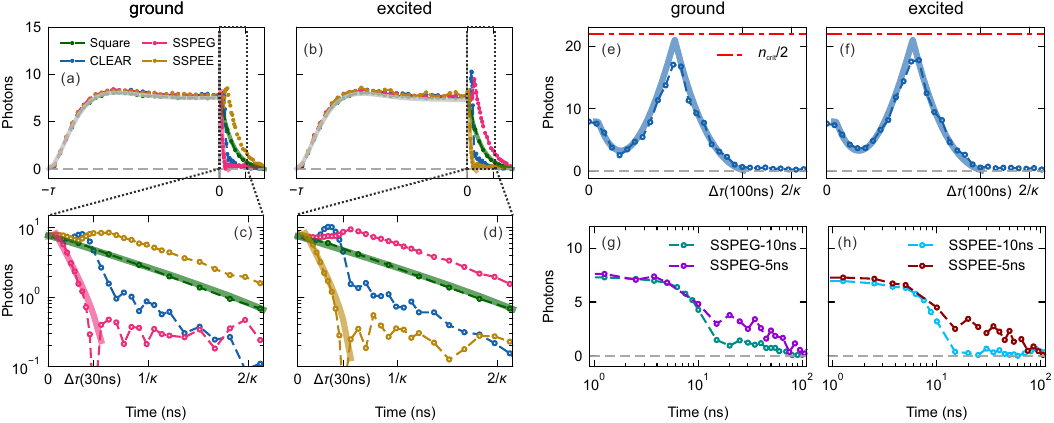}
    \caption{\textbf{Cavity Photon Number Dynamics.} \textbf{(a), (b)} Time-resolved cavity photon number for the qubit prepared in $\ket{g}$ and $\ket{e}$, respectively. The legend labels ``SSPEG" and ``SSPEE" denote the SSPE protocol with pulse parameters configured for the qubit in $\ket{g}$ and $\ket{e}$, respectively. Light gray solid curves denote fits derived from the cavity dynamics model of Eq.~\eqref{eq-1}. The time $t = 0$ corresponds to the onset of the reset segment. \textbf{(c), (d)} Magnified views of the regions enclosed by the dashed boxes in panels (a) and (b), displayed with logarithmic vertical axes. The reset segment duration is fixed at $\Delta \tau = 30$~ns for panels (a)--(d). \textbf{(e), (f)} Cavity photon number dynamics during the reset stage for the CLEAR protocol with $\Delta \tau = 100$~ns for the qubit initialized in $\ket{g}$ and $\ket{e}$, respectively. Light blue solid curves denote numerical solutions obtained from Eq.~\eqref{eq-1}. Red dash-dotted line denotes cavity photon number of $n_{\text{crit}}/2$. Legend conventions follow those defined in panel~(a). \textbf{(g), (h)} Cavity photon number dynamics during the reset stage for the SSPE protocol with reduced reset segment durations of 10~ns and 5~ns, for the qubit initialized in $\ket{g}$ and $\ket{e}$, respectively.}
    \label{fig-fig5}
\end{figure*}

\section{Cavity Photon Number Dynamics}
\label{sec: appendixB}

As depicted in Fig.~\ref{fig-fig2}(a), the cavity photon number dynamics are measured by sweeping the delay and frequency of the spectroscopic $\pi$-pulse. The complete temporal evolution of the combined readout and reset processes for the four different pulses are presented in Fig.~\ref{fig-fig5}(a) and~\ref{fig-fig5}(b) for the qubit initialized in $\ket{g}$ and $\ket{e}$, respectively. For the SSPE protocol, an incorrect initial qubit state not only renders the cavity reset ineffective but also induces a significant photon number overshoot. As illustrated in the magnified views of the reset stage dynamics in Fig.~\ref{fig-fig5}(c) and~\ref{fig-fig5}(d), this overshoot results in a subsequent decay rate that is appreciably slower than passive free decay. 

The reset dynamics for shorter durations of 10~ns and 5~ns are illustrated in Fig.~\ref{fig-fig5}(g) and~\ref{fig-fig5}(h). While the 10-ns case exhibits a noticeable degradation in reset performance, the 5-ns case results in an almost complete failure of the reset protocol. As discussed in the main text, this degradation is attributed to the finite analog bandwidth of the measurement chain, which distorts the rapidly varying transient signals.

\begin{figure}
    \centering
    \includegraphics[width=\columnwidth]{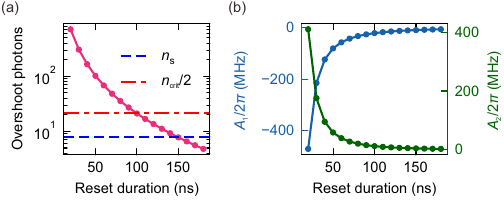}
    \caption{\textbf{Simulated Cavity Photon Number Overshoot of CLEAR Protocol.} \textbf{(a)} Peak photon number overshoot as a function of reset duration, with both segment durations taken to be equal. Red dash-dotted and blue dashed horizontal lines indicate $n_{\text{crit}}/2$ and the steady photon number $n_\text{s}$, respectively. \textbf{(b)} First reset segment amplitude $A_1$ (blue, left vertical axis) and second reset segment amplitude $A_2$ (dark green, right vertical axis) as functions of the reset duration.}
    \label{fig-fig6}
\end{figure}

As shown in Fig.~\ref{fig-fig5}(e) and \ref{fig-fig5}(f), the CLEAR protocol achieves complete cavity photon depletion at a reset duration of 100~ns. To elucidate the physical origin of the photon number overshoot, Eq.~\ref{eq-1} is solved numerically for the CLEAR pulse. The overshoot photon number, defined here as the cavity photon number at the end of the first reset segment, exhibits a super-exponential decrease with increasing reset segment duration, as shown in Fig.~\ref{fig-fig6}(a). The simulated cavity photon number dynamics for a reset duration of 100~ns are illustrated by the light blue curves in Fig.~\ref{fig-fig5}(e) and \ref{fig-fig5}(f), showing excellent agreement with the experimental results. Correspondingly, Fig.~\ref{fig-fig6}(b) shows that the drive amplitudes of both reset segments, $A_1$ and $A_2$, decrease sharply with increasing reset duration. Notably, however, the peak photon number overshoot observed under the 100-ns reset condition exceeds that measured at 30~ns, shown in Fig.~\ref{fig-fig5}(c) and~\ref{fig-fig5}(d). This apparent discrepancy arises because, in the 30~ns case, the rapid transient photon number overshoot is temporally averaged over the finite duration (30~ns) of the $\pi$-pulse, rendering the extracted photon number an underestimate of the true peak value. This constitutes a systematic measurement error inherent to the ac-Stark-shift-based characterization method employed in the present work.

\begin{figure}
    \centering
    \includegraphics[width=0.9\columnwidth]{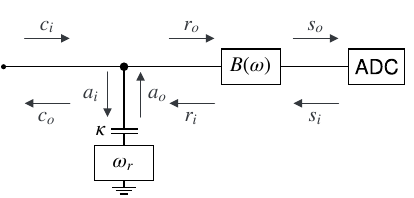}
    \caption{\textbf{Schematic of the Readout Circuit.} The input–output network includes the feedline, readout resonator (\( \omega_r \)), their capacitive coupling characterized by the resonator decay rate $\kappa$, background response \(B(\omega)\), acquisition stage (ADC), and the associated traveling modes (arrows).}
    \label{fig-fig7}
\end{figure}

\begin{figure}
    \centering
    \includegraphics[width=\columnwidth]{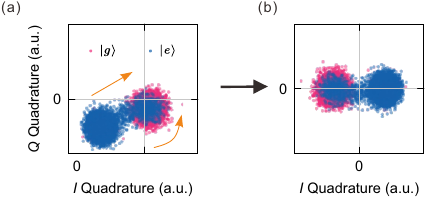}
    \caption{\textbf{IQ clusters Components Calibration.} \textbf{(a)} Raw single-shot readout IQ clusters acquired over $10^4$ repetitions. \textbf{(b)} The Rotation- and translation-calibrated single-shot IQ clusters.}
    \label{fig-fig8}
\end{figure}

\begin{figure*}
    \centering
    \includegraphics[width=2\columnwidth]{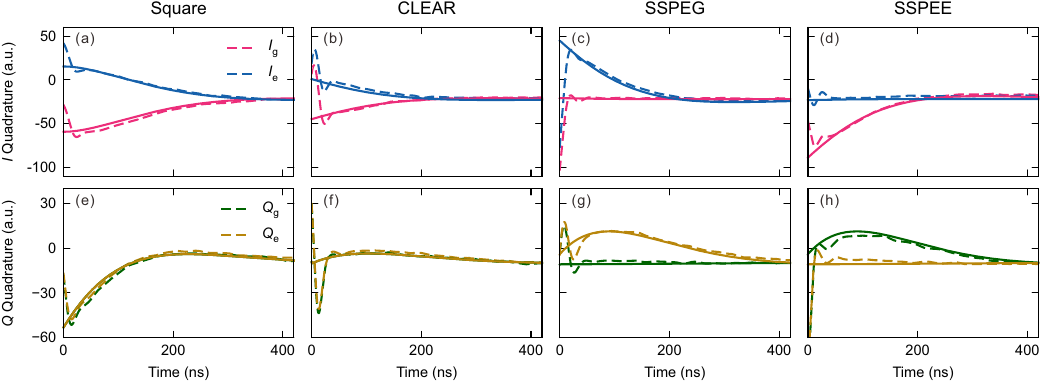}
    \caption{\textbf{Calibrated Time-Resolved Output Field IQ Trajectories.} 
    Pink (dark green) and blue (gold) curves refer to I(Q) quadratures for qubit states $\ket{g}$ and $\ket{e}$, respectively. Dashed and solid curves represent the calibrated ADC data and numerical simulations, respectively.
    Panels \textbf{(a), (e)} Square pulse. \textbf{(b), (f)} CLEAR pulse. \textbf{(c), (g)} SSPE protocol with pulse parameters configured for $\ket{g}$. \textbf{(d), (h)} SSPE protocol with pulse parameters configured for $\ket{e}$.
    }
    \label{fig-fig9}
\end{figure*}

\section{Calibration of Output Field IQ Trajectories}
\label{sec: appendixC}
In the readout circuit, the resonator is capacitively coupled to the feedline through a three-port junction, as illustrated in Fig.~\ref{fig-fig7}. The incoming (\(i\)) and outgoing (\(o\)) traveling modes at the input port are denoted by \(c_i\) and \(c_o\), those at the resonator port by \(a_i\) and \(a_o\), and those at the output port by \(r_i\) and \(r_o\), respectively. The measurement chain introduces a nontrivial background response, \(B(\omega)\), which incorporates frequency-dependent gain or attenuation as well as phase delay. For simplicity, these effects are lumped into an effective background response immediately preceding the acquisition stage. The resulting signal is subsequently recorded by the analog-to-digital converter (ADC), with the corresponding traveling modes denoted by \(s_i\) and \(s_o\).

The input-output relation at the resonator port is

\begin{equation}
    a_o = a_i - \sqrt{\kappa} a
    \label{eq-c1}
\end{equation}
where \(a\) denotes the intracavity field mode. The scattering relations for an ideal symmetric three-port junction can be written as~\cite{dedicated_filter},

\begin{equation}
    c_o = -\frac{1}{3} c_i + \frac{2}{3} r_i + \frac{2}{3} a_o,
    \label{eq-c2}
\end{equation}

\begin{equation}
    r_o = \frac{2}{3} c_i - \frac{1}{3} r_i + \frac{2}{3} a_o,
    \label{eq-c3}
\end{equation}

\begin{equation}
    a_i = \frac{2}{3} c_i + \frac{2}{3} r_i - \frac{1}{3} a_o,
    \label{eq-c4}
\end{equation}

Combining Eqs.~\ref{eq-c1}, \ref{eq-c3}, and \ref{eq-c4} and eliminating the intermediate traveling modes \(a_i\) and \(a_o\), we obtain

\begin{equation}
    r_o = c_i - \frac{\sqrt{\kappa}}{2} a.
    \label{eq-c5}
\end{equation}

Equation~\ref{eq-c5} explicitly shows that, while the resonator is being driven, the output field generally contains contributions from both the directly transmitted input field and the field emitted by the resonator. Once the external drive pulse is terminated, however, the input field vanishes, \(c_i=0\), and Eq.~\ref{eq-c5} reduces to

\begin{equation}
    r_o = -\frac{\sqrt{\kappa}}{2}a.
    \label{eq-c6}
\end{equation}

Thus, after termination of the external drive, the output field is directly proportional to the residual intracavity coherent field. Before reaching the ADC, the output field propagates through the amplification, downconversion, and acquisition chain. The acquired field can be expressed as,

\begin{equation}
    s_o
    = B(\omega) r_o
    = -\frac{\sqrt{\kappa}}{2}B(\omega)a,
    \quad c_i=0,
    \label{eq-c7}
\end{equation}
where \(B(\omega)\) denotes the effective background response of the measurement chain. An additional calibration removes the associated gain or attenuation, phase delay, and offset but does not reconstruct the intracavity field during the driven interval. Rather, once the drive is switched off, Eq.~\ref{eq-c6} establishes a direct proportionality between \(r_o\) and \(a\), so the calibrated ADC field signal reproduces the intracavity field trajectory up to an overall scale factor.

The calibration procedure is illustrated in Fig.~\ref{fig-fig8}. As shown in Fig.~\ref{fig-fig8}(a), the raw IQ clusters are subjected to a rotation about their centroid followed by a translation to the origin. The resulting distributions, shown in Fig.~\ref{fig-fig8}(b), are symmetric with respect to the origin, ensuring that the state information is concentrated in the I quadrature while the Q quadrature exhibits no discernible state-dependent difference.

The calibrated time-resolved output field IQ trajectories are shown in Fig.~\ref{fig-fig9}. The time origin, \(t=0\), is redefined as the end of the complete readout pulse, including the reset segment. Consequently, for \(t\geq0\), the external input field is absent and the signal recorded by the ADC originates solely from the residual field emitted by the resonator. The dashed curves in Fig.~\ref{fig-fig9} represent the calibrated ADC data, whereas the solid curves are obtained by numerically solving Eq.~\ref{eq-1}. For comparison with the measured signal, the numerical intracavity field trajectories are transformed using a single global scaling factor that accounts for the overall gain of the measurement chain, together with a constant offset that accounts for the zero-point shift associated with the IQ calibration procedure.

For the SSPE protocol, the output field successfully returns to the vacuum state upon completion of the pulse, as shown in Figs.~\ref{fig-fig9}(c, g, d, h). In contrast, when the reset parameters are mismatched with the actual initial qubit state, a finite residual field remains. The CLEAR protocol [Fig.~\ref{fig-fig9}(b) and \ref{fig-fig9}(f)] leaves a residual field at the end of the pulse, which subsequently decays passively. For the Square pulse, the output field likewise undergoes slow passive decay. These observations are consistent with the corresponding photon number dynamics following the $\Delta \tau$ reset segment shown in Fig.~\ref{fig-fig5}(c) and \ref{fig-fig5}(d).

\section{Qubit-State Dependence and LAFP Scaling of the SSPE Pulse.}
\label{sec: appendixD}
Here, we derive the qubit-state dependence and the LAFP scaling property of the SSPE pulse within the linear dispersive model of Eq.~\ref{eq-1}. To achieve cavity reset, the boundary condition requiring the intracavity field amplitude to vanish at the end of the reset segment ($t = \Delta\tau$) must be satisfied. Imposing $\alpha(\Delta\tau) = 0$ in Eq.~\ref{eq-5} yields

\begin{equation}
    \varepsilon_r e^{i\phi_r} \left(1-e^{\frac{1}{2} \Delta \tau C_j}\right) - \varepsilon_n \left(1-e^{-\frac{1}{2} \tau C_j}\right) = 0 + i 0,
    \label{eq-D1}
\end{equation}
where $C_j = 2i(\Delta_r + \chi_j) + \kappa$. Solving Eq.~\ref{eq-D1} gives the optimal complex reset vectors for the qubit state $\ket{j}$,

\begin{equation}
\frac{z_{r,j}}{\varepsilon_n}
=
\frac{\varepsilon_{r,j}e^{i\phi_{r,j}}}{\varepsilon_n}
=
\frac{1-e^{-\frac{1}{2}\tau C_j}}
{1-e^{\frac{1}{2}\Delta\tau C_j}}
\equiv
F_j
=
\Lambda_j e^{i\Psi_j},
\label{eq-D2}
\end{equation}
where $\Lambda_j=|F_j|$ and $\Psi_j=\arg(F_j)$. The coefficient $F_j$ depends on the qubit state through the dispersive shift $\chi_j$. Consequently, the optimal reset vectors for $\ket{g}$ and $\ket{e}$ are generally different. A single complex-constant reset segment can deplete the cavity for both states only if 
\begin{equation}
F_g=F_e,
\label{eq-D3}
\end{equation}

This condition is not satisfied for the experimental parameters considered here. It is trivially recovered in the degenerate limit $\chi\to0$, where the two qubit states become indistinguishable and the readout itself fails.

Although the optimal amplitudes may be identical, the different optimal phases imply \(z_{r,g}\neq z_{r,e}\). Therefore, a reset vector calibrated for one qubit state generally leaves a finite residual cavity field when applied to the other state, thereby establishing the qubit-state-dependent limitation of the present SSPE protocol. This suggests that a state-independent reset generally requires at least a second complex pulse segment, as in CLEAR~\cite{CLEAR}, or a continuously time-varying amplitude and/or phase envelope.

For fixed pulse parameters $\tau$ and $\Delta\tau$, the coefficient $F_j$ is independent of the readout amplitude. Separating its magnitude and phase gives
\begin{equation}
\frac{\varepsilon_{r,j}}{\varepsilon_n}
=
\Lambda_j,
\qquad
\phi_{r,j}
=
\Psi_j+2n\pi,
\quad
n\in\mathbb{Z}.
\label{eq-D4}
\end{equation}

Equation~\ref{eq-D4} explicitly demonstrates that the optimal reset amplitude scales proportionally with the readout amplitude, whereas the optimal reset phase remains strictly invariant under such rescaling. The corresponding amplitude scaling ratios and optimal phases for the various reset durations are summarized in Table~\ref{tab:LAFP_value}. At a reset duration of 10~ns, the discrepancy of the optimized reset amplitude parameters is mainly attributed to the finite analog bandwidth of the measurement chain, and consequently shifts the optimal amplitudes associated with the $|g\rangle$ and $|e\rangle$ states away from their ideal values. 

\begin{table}
    \renewcommand\tabcolsep{12pt}
    \centering
    \caption{Reset amplitude ratios $\Lambda$ and reset phase $\Psi$} 
    \begin{tabular}{c|cccc}
    \hline
    \hline
    \\
     Reset Duration & $\Lambda_g$ & $\Psi_g$ & $\Lambda_e$ & $\Psi_e$ \\
    \\
    \hline
    100~ns          & 0.6 & 2.10  & 0.6 & 4.12  \\
    50~ns           & 1.6 & 2.31  & 1.6 & 4.08  \\
    30~ns           & 2.8 & 2.40  & 2.8 & 3.89  \\
    10~ns           & 7.8 & 2.47  & 8.8 & 3.89  \\
    \hline
    \hline
    \end{tabular}
    \label{tab:LAFP_value}
\end{table}

\section{Non-QNDness Measurement.}
\label{sec: appendixE}

\begin{figure}
    \centering
    \includegraphics[width=\columnwidth]{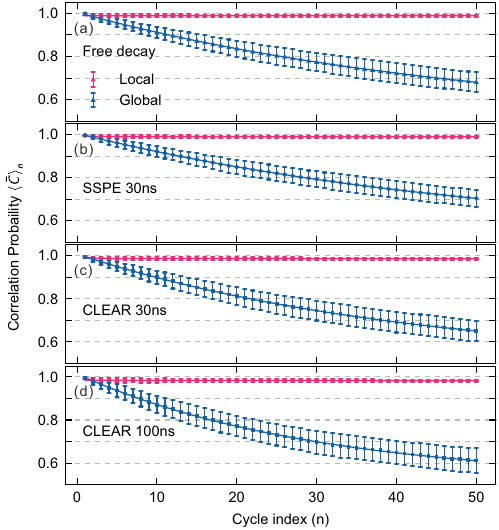}
    \caption{\textbf{QNDness Characterization Across Different Reset Configurations.} Average local (pink) and global (blue) correlation probabilities as a function of the measurement cycle index $n$ for the \textbf{(a)} free decay, \textbf{(b)} 30-ns SSPE, \textbf{(c)} 30-ns CLEAR, and \textbf{(d)} 100-ns CLEAR protocols.}
    \label{fig-fig10}
\end{figure}

As illustrated in Fig.~\ref{fig-fig4}(a), the RILRB experimental protocol comprises $n$ repeated readout operations interleaved with randomized qubit flips. The initial readout serves as a pre-selection measurement implemented by the SSPE reset pulse that depletes the cavity population, and only experimental instances in which the qubit is measured into the ground state are retained through post-selection. Each combined measurement-and-reset (MR) block incorporates a reset component with a duration of 400~ns. This extended interval guarantees the complete dissipation of cavity photons prior to the subsequent operation, even in the event of faulty state preparation within the SSPE scheme caused by leakage or state-switching errors. During a fixed time window $\Delta \tau$ at the onset of this reset component, various active reset protocols are implemented. Crucially, for the SSPE protocol, the reset parameters are deterministically conditioned on the expected qubit state, which is pre-calculated from the cumulative control sequence applied prior to each measurement.

The outcome of the $n$-th measurement is denoted $r_n \in \{0, 1\}$, and the stochastic operation preceding each readout is defined as $i_n \in \{0, 1\}$, where $i_n = 1$ corresponds to a $\pi$-pulse ($X$ gate) and $i_n = 0$ corresponds to the identity operation ($I$ gate). 

To quantify the local correlation between successive measurements, we compute the binary ``flipped-or-not" quantity, defined as the XOR sum of consecutive outcomes:

\begin{equation}
    o_n = r_{n-1} \oplus r_n 
\end{equation}
and compare it to the intended operation $i_n$ to obtain the local correlation metric:
\begin{equation}
    C_n^{\text{local}}=1 - o_n \oplus i_n
\end{equation}
in this framework, $C_n^{\text{local}} = 1$ signifies that the measurement outcome is logically consistent with the applied gate operation. 

To evaluate the global correlation, we compare the qubit state after $n$ cycles to the initial heralded ground state $\ket{g}$, define the accumulated operation quantity:
\begin{equation}
    h_n = i_1 \oplus i_2 \oplus \cdots \oplus i_n
\end{equation}
from which the global correlation metric is defined as:

\begin{equation}
    C_n^{\text{global}}=1 - r_n \oplus h_n
\end{equation}
in this convention, $C_n^{\text{global}} = 1$ indicates that the measured outcome $r_n$ is perfectly consistent with the net effect of the applied gate sequence.

To determine the average correlation $\bar{C}_n$, we perform $m$ experimental realizations for each unique random sequence and subsequently average over different randomizations to obtain the ensemble mean correlation probability $\langle \bar{C}_n \rangle$. In the present experiments, $m = 10^4$ realizations are performed per sequence, averaged over 300 independent random sequences.

Following Ref.~\cite{2025_IQM_high_Single_Two_Readout}, the local and global correlations evolve with the cycle $n$ as:
\begin{equation}
    \langle \bar{C}_n^{\text{local}} \rangle = \frac{L(A-\frac{1}{2})}{L+S} \left( 1-L-S \right)^n + \frac{AS+\frac{1}{2}L}{L+S},
\end{equation}

\begin{equation}
    \langle \bar{C}_n^{\text{global}} \rangle = \left( \frac{1}{2} - e \right) (1-2h) (1-2p-L)^n +  \frac{1}{2},
\end{equation}
where $L=\Sigma_i^k L_i, S=\Sigma_i^k S_i$ denote the average leakage and seepage probabilities per readout, respectively. Specifically, $L_i = \left( L_{g,i} + L_{e,i} \right) /2$ is the average leakage rate from computational subspace to leakage state $\ket{l_i}$, and $S_i = \left( S_{g,i} + S_{e,i} \right) /2$ is the average seepage rate from leakage state $\ket{l_i}$ back to computational subspace. The quantity $p=(p_e + p_g) /2$ denotes average state switch rate. Readout and state-initialization errors are accounted for by the average readout error $e = (e_0 + e_1)/2$ and the initialization error $h$. The coefficient $A$ depends on all other model parameters and is given by

\begin{equation}
    A=\frac{1}{2} \sum_{s_n \in \lbrace g,e \rbrace} \sum_{s_{n+1}} P\left( r_{n-1} \oplus r_n \oplus i_{n+1} =0 ,s_{n+1} | s_{n}\right),
\end{equation}
where $s_n$ denotes the physical qubit state following the $n$-th measurement. The conditional mapping $P(a, b|c)$, denotes the probability that the qubit ends up in state $a$ and is assigned outcome $b$, given that the qubit is initially in state $c$.
The individual transition probabilities are defined as follows:
\begin{align}
    L_{g,i} &= P\left(l_i, 0| g\right) + P\left(l_i, 1| g\right) \nonumber \\
    L_{e,i} &= P\left(l_i, 0| e\right) + P\left(l_i, 1| e\right) \nonumber\\
    S_{g,i} &= P\left(g, 0| l_i\right) + P\left(g, 1| l_i\right) \nonumber\\
    S_{e,i} &= P\left(e, 0| l_i\right) + P\left(e, 1| l_i\right) \\
    p_g &= P\left(e, 0 | g \right) + P\left(e, 1 | g \right) \nonumber \\
    p_e &= P\left(g, 0 | e \right) + P\left(g, 1 | e \right) \nonumber \\
    e_0 &= P\left( 0| e \right) \nonumber\\
    e_1 &= P\left( 1| g \right) \nonumber
\end{align}

The leakage rate $L$ and seepage rate $S$ are estimated by fitting the amplitude, offset, and decay rate of the exponentially decaying correlation data. In the limiting case where only a single leakage state is relevant, the local correlation provides a direct measure of $L$. Combining this with the global correlation decay constant, the QND fidelity is determined as

\begin{equation}
    Q = 1 - p - L,
\end{equation}

The corresponding experimental results for passive free decay, 30-ns SSPE, 30-ns CLEAR and 100-ns CLEAR protocols are presented in Fig.~\ref{fig-fig10}. 

\bibliography{ref.bib}

\end{document}